\documentclass[journal,11pt,onecolumn,draftclsnofoot,]{IEEEtran}

\usepackage{graphicx}                % For including figures
\usepackage{amsmath}                 % For advanced math formatting
\usepackage{amsfonts}                % For additional math fonts
\usepackage{amssymb}                 % For more math symbols
\usepackage{cite}                    % For managing citations
\usepackage{hyperref}                % For hyperlinks in the document
\usepackage{tabularx}                % For flexible table formatting
\usepackage{multirow}                % For multi-row cells
\usepackage{booktabs}                % For professional table lines
\usepackage{float}                   % For [H] placement option
\usepackage{array}
\usepackage{xcolor} % Optional for cell shading
\usepackage{adjustbox} % For table adjustments
\usepackage{wrapfig} % For wrapping figures or tables

\usepackage{float}
\usepackage{makecell}
\usepackage{algorithm}
\usepackage{algorithmic}
\usepackage[USenglish]{babel}
\usepackage[utf8]{inputenc}
\usepackage{rotating}
\usepackage{array,amsmath,graphicx}

\usepackage{pdflscape}
\usepackage{afterpage}
\usepackage{capt-of}% or use the larger `caption` package

\usepackage{lipsum}% dummy tex
\usepackage{cite}
\usepackage{csquotes}

\def\tsc#1{\csdef{#1}{\textsc{\lowercase{#1}}\xspace}}
\tsc{WGM}
\tsc{QE}
\tsc{EP}
\tsc{PMS}
\tsc{BEC}
\tsc{DE}
\usepackage{framed}

\usepackage{bm}

\usepackage{nomencl}
\usepackage{bbm}

\renewcommand\nomgroup[1]{%
    \item[\bfseries
     \ifstrequal{#1}{O}{\textbf{Operators}}{%
        \ifstrequal{#1}{S}{\textbf{Indices and sets}}{%
            \ifstrequal{#1}{V}{\textbf{Variables and parameters}}{%
            }}}
            ]}
            
            \makenomenclature

\makeatletter
\newsavebox\myboxA
\newsavebox\myboxB
\newlength\mylenA

\newcommand*\xoverline[2][0.75]{%
    \sbox{\myboxA}{$\m@th#2$}%
    \setbox\myboxB\null% Phantom box
    \ht\myboxB=\ht\myboxA%
    \dp\myboxB=\dp\myboxA%
    \wd\myboxB=#1\wd\myboxA% Scale phantom
    \sbox\myboxB{$\m@th\overline{\copy\myboxB}$}%  Overlined phantom
    \setlength\mylenA{\the\wd\myboxA}%   calc width diff
    \addtolength\mylenA{-\the\wd\myboxB}%
    \ifdim\wd\myboxB<\wd\myboxA%
      \rlap{\hskip 0.5\mylenA\usebox\myboxB}{\usebox\myboxA}%
    \else
        \hskip -0.5\mylenA\rlap{\usebox\myboxA}{\hskip 0.5\mylenA\usebox\myboxB}%
    \fi}
\makeatother
\makenomenclature

\usepackage{subcaption}
\usepackage{mleftright}
\usepackage{hyperref}

\usepackage{amsmath}
\usepackage{amssymb}
\usepackage{amsfonts}
\usepackage{pgfplotstable}
\usetikzlibrary{spy}
\usetikzlibrary{calc}
\usetikzlibrary{fadings}
\usetikzlibrary{patterns}
\usetikzlibrary{shadows}
\usetikzlibrary{mindmap}
\usetikzlibrary{backgrounds}
\usetikzlibrary{shapes.symbols}
\usetikzlibrary{shapes.multipart}
\usetikzlibrary{shapes.geometric}
\usetikzlibrary{automata,positioning}
\usetikzlibrary{decorations.fractals} 
\usetikzlibrary{decorations.markings}
\usetikzlibrary{decorations.pathreplacing}
\usetikzlibrary{decorations.pathmorphing}
 
\usepackage{bm}

\usepackage{nomencl}
\makenomenclature
\tikzset{edge from parent/.style={segment angle=10,draw}}

\tikzset{
 my rounded corners/.append style={rounded corners=2pt},
}

\def\BibTeX{{\rm B\kern-.05em{\sc i\kern-.025em b}\kern-.08em
 T\kern-.1667em\lower.7ex\hbox{E}\kern-.125emX}}

\renewcommand{\nomgroup}[1]{%
 \ifthenelse{\equal{#1}{O}}{\item[\textit{Operators}]}{%
 \ifthenelse{\equal{#1}{I}}{\item[\textit{Indices}]}{%
 \ifthenelse{\equal{#1}{A}}{\item[\textit{Acronyms}]}{%
 `\ifthenelse{\equal{#1}{V}}{\item[\textit{Variables and parameters}]}{}}}}}
\usetikzlibrary{svg.path}
\definecolor{orcidlogocol}{HTML}{A6CE39}
\tikzset{
 orcidlogo/.pic={
 \fill[orcidlogocol] svg{M256,128c0,70.7-57.3,128-128,128C57.3,256,0,198.7,0,128C0,57.3,57.3,0,128,0C198.7,0,256,57.3,256,128z};
 \fill[white] svg{M86.3,186.2H70.9V79.1h15.4v48.4V186.2z}
 svg{M108.9,79.1h41.6c39.6,0,57,28.3,57,53.6c0,27.5-21.5,53.6-56.8,53.6h-41.8V79.1z M124.3,172.4h24.5c34.9,0,42.9-26.5,42.9-39.7c0-21.5-13.7-39.7-43.7-39.7h-23.7V172.4z}
 svg{M88.7,56.8c0,5.5-4.5,10.1-10.1,10.1c-5.6,0-10.1-4.6-10.1-10.1c0-5.6,4.5-10.1,10.1-10.1C84.2,46.7,88.7,51.3,88.7,56.8z};
 }
}

\newcommand\orcidicon[1]{\href{https://orcid.org/#1}{\mbox{\scalerel*{ \begin{tikzpicture}[yscale=-1,transform shape]
 \pic{orcidlogo};
 \end{tikzpicture}
 }{|}}}}

\begin{document}

\title{Fairness for distribution network hosting capacity }
\author{Olivia Rubbers$^{1}$, Sari Kerckhove$^{1}$, 
~Md~Umar~Hashmi$^{1}$,~and~Dirk~Van~Hertem$^{1}$
\thanks{Corresponding author email: mdumar.hashmi@kuleuven.be}
\thanks{$^{1}$O.R., S.K., M.U.H. and D.V.H. are with KU Leuven \& EnergyVille, Belgium.}
\thanks{This work is supported by the Flemish Government and Flanders Innovation \& Entrepreneurship (VLAIO) through the Flux50 project IMPROcap (HBC 2022.0733) {and KU Leuven funded C2 project FlexIQ}.}}

\IEEEoverridecommandlockouts
\IEEEpubid{\makebox[\columnwidth]{979-8-3503-9042-1/24/\$31.00 \copyright 2025 IEEE \hfill } 
\hspace{\columnsep}\makebox[\columnwidth]{\hfill }}

% \author{\IEEEauthorblockN{Md Umar Hashmi, Arpan Koirala, Hakan Ergun, Dirk Van Hertem}
% \IEEEauthorblockA{\textit{Electa-ESAT, KU Leuven \& EnergyVille},
% % \textit{name of organization (of Aff.)}\\
% Genk, Belgium \\
% (mdumar.hashmi, arpan.koirala, hakan.ergun, dirk.vanhertem)@kuleuven.be}
% \and
% \IEEEauthorblockN{Arpan Koirala}
% \and
% \IEEEauthorblockN{Hakan Ergun, Dirk Van Hertem}
% \and
% \and
% \and

\maketitle

\begin{abstract}
The integration of distributed generation (DG) is essential to the energy transition but poses challenges for low-voltage (LV) distribution networks (DNs) with limited hosting capacity (HC). This study incorporates multiple fairness criteria, utilitarian, egalitarian, bounded, and bargaining, into the HC optimisation framework to assess their impact. When applied to LV feeders of different sizes and topologies, the analysis shows that bargaining and upper-bounded fairness provide the best balance between efficiency and fairness. 
Efficiency refers to maximising the social welfare of the LV DNs, while fairness is proportional to the minimisation of disparity in opportunity for installing DG.
Feeder topology significantly influences fairness outcomes, while feeder size affects total HC and the inherent fairness of feeders. These results emphasise the importance of regulatory incentives and network designs in order to facilitate fair and efficient DG integration.

\end{abstract}

\vspace{20pt}
{\textbf{Disclaimer}: This paper is a preprint of a paper submitted to and to be presented at the IEEE ISGT Europe 2025. 
The final version of the paper will be available at IEEE xplore.}
% If accepted, the copy of record will be available at IET Digital Library}

\pagebreak

\tableofcontents

\pagebreak

\section{Introduction}
\label{chap1: introduction}
% \subsection{Motivation and background}
With the growing integration of distributed generation (DG) driven by the energy transition, low-voltage (LV) distribution networks (DN) are increasingly exposed to variable and bidirectional power flows. This shift poses significant challenges, as it may result in the violation of technical constraints and lead to grid congestion. To ensure secure and reliable grid operation, it is essential to assess the hosting capacity (HC) of LV networks, quantifying the maximum DG that can be accommodated without breaching operational limits. Beyond technical considerations, fairness has emerged as a critical dimension in HC assessments. Given that the structure and topology of LV networks inherently lead to unequal access to grid capacity among users, incorporating fairness into HC evaluations is necessary to promote equitable access and avoid the systematic 
% exclusion of 
disparity of opportunity to integrate DG for
certain customers.

\subsection{Relevant literature and Research gap}
HC and its assessment methodologies in LV distribution networks have been extensively studied to determine the maximum allowed integration of DG, such as PV systems, without violating grid constraints \cite{gupta2024analysis, hashmi2025hosting, ismael2019state, ramesh2021fair, singh2023implementing, zain2020review}. For instance, Ismael et al. provide a comprehensive review of HC definitions and enhancement strategies, while Zain et al. examine various tools and methodologies for HC calculations \cite{ismael2019state, zain2020review}. The influence of network topology and structural parameters on HC has also been studied \cite{puvi2023evaluating, lin2024study}. For example, Puvi et al. investigate the impact of MV and LV network size and configuration on HC in high PV penetration scenarios, from both the DSO and customer perspectives.

Fairness in DNs, particularly in optimisation contexts, has been formalised through a variety of mathematical frameworks. Xinying et al., provide a detailed overview of equitable utility allocation strategies and propose several inequality metrics \cite{xinying2023guide}. The explicit integration of fairness into HC analysis, however, is a relatively recent development. Aydin et al. incorporate fairness by allocating HC proportionally to each consumer’s load, while Mahmoodi et al. define fairness as an equal distribution of DG capacity among all consumers, reflecting an egalitarian approach \cite{aydin2025fairness, mahmoodi2021hosting}.
Similarly, Fani et. al. utilizes equitable electric vehicle charging placement HC \cite{fani2024impact}.
Prior works, such as \cite{koirala2022decoupled}, consider a utilitarian approach for DG HC assessment.

In situations where HC is limited and congestion emerges, fairness has predominantly been applied in the context of reactive or retrospective PV curtailment strategies. These studies typically address dynamic control problems rather than planning approaches such as the one considered in this study. For instance, Ahmed et al. and Gerdroodbari et al. propose algorithms to fairly distribute active power curtailment based on users’ spatial positions along the feeder, ensuring no user is disproportionately impacted \cite{ahmed2024equitable, gerdroodbari2021decentralized}. Similarly, Alyami et al. introduce an adaptive real power capping method to fairly regulate overvoltages in PV-rich networks, balancing HC utilisation with equitable prosumer impact \cite{alyami2014adaptive}. 
% To avoid curtailment altogether, researchers have explored HC enhancement measures such as smart control, network reconfiguration, and optimal DG placement. These approaches not only increase HC but also support equitable access to network capacity \cite{kerckhove2023reconfiguration, qamar2023hosting}. 
Kerckhove et al. in \cite{kerckhove2023reconfiguration} demonstrate how grid reconfiguration can improve fair HC by optimising DG placement. Qamar et al. in \cite{qamar2023hosting} further highlight the role of topological design and enhancement techniques in maximising HC and mitigating inequality.

% \subsection{Research gap}
% Despite HC advancements, fairness in HC allocation is often invoked without a rigorous definition or consistent application. Existing studies tend to employ a single fairness concept, often failing to quantify the trade-offs between fairness and overall utility. To date, no study has systematically compared multiple fairness definitions in the context of HC, nor evaluated them using formal inequality and efficiency metrics. Furthermore, the impact of network topology on the balance between fairness and efficiency in HC distribution remains insufficiently explored. This research addresses these gaps by conducting a quantitative analysis of various fairness criteria in HC allocation and examining how they perform under different topological conditions.
\textbf{Research gap}:
Despite advances in HC, fairness is often vaguely defined and inconsistently applied. Most studies use a single fairness concept without quantifying trade-offs with overall utility. No prior work has systematically compared multiple fairness definitions in HC or assessed them using formal inequality and efficiency metrics.
Furthermore, the impact of network topology on the balance between fairness and efficiency in HC distribution remains insufficiently explored. This research addresses these gaps by conducting a quantitative analysis of various fairness criteria in HC allocation and examining how they perform under different topological conditions.
% The role of network topology in balancing fairness and efficiency also remains underexplored. This study addresses these gaps through a quantitative analysis of fairness criteria and their performance across different topologies.

% \vspace{-10pt}

\subsection{Contributions and organization}
% This paper addresses this gap by proposing several definitions of fairness for the DG HC optimisation-based framework and comparing the outcomes of these different implementations. 
% Furthermore, this work also examines the impact of DN size and topology on DG HC and establishes a framework for improving HC.
We propose the notions of fairness for the DG HC problem, considering the system operator and the electricity consumer as stakeholders.
The fairness-aware optimisation-based DG HC frameworks are implemented and compared for four representative feeders.
This work also analyses how DN size and topology affect fair and efficient DG HC.
% We propose fairness concepts for the DG HC problem, considering system operators and consumers. Fairness-aware, optimisation-based DG HC frameworks are implemented and compared across four representative feeders. The impact of DN size and topology on fair and efficient DG HC is also analysed.
% and compared with the outcomes of these different implementations.
% and proposes a framework to enhance it.

This paper is organised into sections.
In Section \ref{chap2: optimisation formulation}, the HC optimisation problem is described and the different fairness definitions are detailed. Section \ref{chap3: KPI} outlines the key performance indicators (KPI) for evaluating the results. Section \ref{chap4: numerical results} presents the numerical results from the different optimisation problems and thoroughly discusses them. Finally, Section \ref{chap5: conclusion} summarises the results and provides useful conclusions.

\pagebreak

\section{Optimisation formulation for HC with fairness}
\label{chap2: optimisation formulation}
% This section presents the PV HC calculation method and the different fairness formulations that are incorporated in the HC assessment.
This section outlines the PV HC calculation method and fairness formulations used in the assessment.

% \subsection{HC calculation method}
The objective of the HC optimisation problem is to maximise the DG HC of the grid. The HC can be expressed as the sum of the maximum allowable active power injections from distributed generation at all load buses, as shown in \eqref{equ: HC max objective}. This objective is the global objective for all the implementations of the HC problem, but for some fairness definitions, it is subject to variations.
\begin{equation}
    HC = \max \sum_{d \in \mathcal{D}} P_{d}^{DG}
    \label{equ: HC max objective}
\end{equation}

The HC optimisation problem is an adaptation of the optimal power flow (OPF) problem. The constraints associated with the OPF problem ensure that HC calculation satisfies the physical laws and power quality constraints.
% needs to be a solution of the deterministic optimal power flow (OPF) problem in order to satisfy the laws of physics. 
Furthermore, the grid must comply with some operational limits to ensure reliable performance. The objective is thus subject to the following constraints:
\begin{equation}
    P_{m} = g V_{m}^{2} - g V_{m} V_{n} \cos(\theta_m - \theta_n) - b V_{m} V_{n} \sin(\theta_m - \theta_n)
    \label{equ: power flow equations P}
\end{equation}
\begin{equation}
    Q_{m} = - g V_{m}^{2} + g V_{m} V_{n} \cos(\theta_m - \theta_n) - b V_{m} V_{n} \sin(\theta_m - \theta_n)
    \label{equ: power flow equations Q}
\end{equation}
\begin{equation}
    \sum_{c\epsilon C_m}^{}P_{c}^{M} +  \sum_{n\epsilon L_m}^{}P_{mn} + P_{d}^{DG} =  P_{d}^{D} \textit{  } \forall d  \in D
    \label{equ: active power balance}
\end{equation}
\begin{equation}
    \sum_{c\epsilon C_m}^{}Q_{c}^{M} +  \sum_{n\epsilon L_m}^{}Q_{mn} + Q_{d}^{DG} =  Q_{d}^{D}  \textit{ } \forall d  \in D
    \label{equ: reactive power balance}
\end{equation}
\begin{equation}
    - P_{c}^{M,max} \leq  P_{c}^{M} \leq  P_{c}^{M,max} \textit{   } \forall c  \in C
    \label{equ: active power exchange transfo}
\end{equation}
\begin{equation}
    - Q_{c}^{M,max} \leq  Q_{c}^{M} \leq  Q_{c}^{M,max} \textit{  } \forall c  \in C
    \label{equ: reactive power exchange transfo}
\end{equation}
\begin{equation}
    (U_{mn} I_{rated, mn})^2 = P_{mn}^2 + Q_{mn}^2 \textit{  } \forall mn  \in L
    \label{equ: thermal limit}
\end{equation}
\begin{equation}
    V_{m}^{min} \leq V_{m} \leq  V_{m}^{max}  \textit{  } \forall m  \in B
    \label{equ: voltage magnitude}
\end{equation}
\begin{equation}
    \Delta\theta_{m}^{min} \leq \theta_m - \theta_n \leq \Delta \theta_{m}^{max}  \textit{  } \forall m  \in B
    \label{equ: voltage angle}
\end{equation}

Equations \eqref{equ: power flow equations P}-\eqref{equ: reactive power balance} represent Ohm’s and Kirchhoff’s laws, forming the power flow equations and hence the physical constraints of the network. The other equations represent operational constraints. Equations \eqref{equ: active power exchange transfo}-\eqref{equ: reactive power exchange transfo} impose limits on the power exchanged with the upstream grid, while equations \eqref{equ: thermal limit} ensure the lines are not overloaded, and \eqref{equ: voltage magnitude}-\eqref{equ: voltage angle} limit the bus voltages.

\subsection{Notion of fairness for HC}
The notions of fairness for HC can have many interpretations. In this work, we consider the traditional DG HC problem with two key stakeholders: (i) the system operator and (ii) the electricity consumers.
Consumers can incur locational disparity due to unequal opportunity to DN design \cite{hashmi2022can}.
Three types of fairness are considered across five distinct definitions. The first is \textbf{utilitarian} fairness, which aims to maximise total utility and thus maximise overall societal welfare. The second is \textbf{egalitarian} fairness, which emphasises fairness for disadvantaged individuals by ensuring equal access to DG opportunities for all consumers. The remaining three definitions represent trade-offs between efficiency and fairness.
In the context of HC, efficiency refers to maximising the overall social welfare of low-voltage DNs, while fairness focuses on minimising disparities in the opportunity to install DG among users.
The other notions of fairness proposed in this work for the DG HC problem are \textbf{lower-bounded} fairness, \textbf{upper-bounded} fairness and \textbf{bargaining} fairness. Each of these introduces constraints or mechanisms to balance fair outcomes with system efficiency. 

In the case of \textit{utilitarian fairness}, no modifications to the base optimisation formulation are required. HC is maximised regardless of how DG capacity is allocated among individual households. The limits on the installed DG are given as
\begin{equation}
    P_i^{DG} \in [0, P_{\max}^{DG}], ~ \forall i \in \mathcal{D}.
\end{equation}
For the utilitarian fairness case, there are no specific limits on specific parts of the DN, rather a uniform limit for all nodes. Since the DN design is radial in nature, and the upper limit is denoted by $P_{\max}^{DG}$ is high enough, the HC outcome would allocate a lot of DG on some nodes compared to others.

To implement \textit{egalitarian fairness}, however, an additional constraint is introduced to ensure equal DG capacity allocation across all load buses. This guarantees uniform access to DG opportunities for all consumers. Egalitarian fairness condition is given as
\begin{equation}
    P^{DG} = P_1^{DG}= P_2^{DG} \dots = P_\mathcal{D}^{DG}.
\end{equation}

For the \textbf{trade-off-based fairness definitions}, the \textit{bounded fairness} approaches are considered first. In these formulations, the DG capacities at the load buses are constrained within specified upper and lower bounds to achieve a more balanced distribution of the total HC. This approach requires two parameters, an upper and a lower bound, which are systematically varied to analyse the impact of stricter or more relaxed fairness constraints on the overall HC. The evolution of these bounds is described in equations \eqref{equ: bounded lower limit} and \eqref{equ: bounded upper limit}, respectively. The corresponding inequality constraints, which enforce the bounded DG capacities, are given in equation \eqref{equ: bounded constraint}.

\begin{equation}
     P_{min}^{DG} \leq P_{i}^{DG} \leq P_{max}^{DG}, ~ \forall i \in \mathcal{D}.
    \label{equ: bounded constraint}
\end{equation}
\begin{equation}
    P_{min}^{DG} = \alpha \cdot P_{egal}^{DG} \text{ for $\alpha$ in [0, 1]}
    \label{equ: bounded lower limit}
\end{equation}
\begin{equation}
    P_{max}^{DG} = P_{egal}^{DG} + \beta \cdot ( \max(P_{uti}^{DG}) - P_{egal}^{DG}) \text{ for $\beta$ in [0, 1]}
    \label{equ: bounded upper limit}
\end{equation}

Lastly, \textit{bargaining fairness} is implemented using an alternative objective function and a parameter, $K$, that varies between $0$ and $1$. This allows the trade-off between efficiency and fairness to be achieved. The new objective is a convex combination of an efficiency term and a disparity penalty, see \eqref{equ: bargaining obj} with the penalty term defined in \eqref{equ: average DG}, \eqref{equ: 1} and \eqref{equ: 2}. As $K$ approaches $1$, the emphasis on disparity diminishes and the objective converges to a formulation based purely on efficiency (utilitarianism). Conversely, lower values of $K$ increase the weight of the disparity term, encouraging a more fair distribution of DG capacity.
\begin{equation}
    P_{avg}^{DG} = \sum_{d \epsilon D} P_{d}^{DG} / \sum_{d \epsilon D} 1
    \label{equ: average DG}
\end{equation}
\begin{equation}
    \Delta P_{max}^{DG} \geq P^{DG}_{d} - P^{DG}_{avg} \textbf{ }\forall d  \in D
    \label{equ: 1}
\end{equation}
\begin{equation}
    \Delta P_{max}^{DG} \geq P^{DG}_{avg} - P^{DG}_{d} \textbf{ }\forall d  \in D
    \label{equ: 2}
\end{equation}
\begin{equation}
    \text{maximise } ( K \cdot \sum_{m\epsilon B} P_{m}^{DG} - (1-K) \Delta P_{max}^{DG} )
    \label{equ: bargaining obj}
\end{equation}

\pagebreak

\section{Key performance indicators}
\label{chap3: KPI}
To evaluate the various fairness implementations of the HC problem, multiple KPI, that quantify the efficiency of the results and the inequality of the distribution are used. The KPI used to quantify the efficiency of the obtained solution is the \textit{price of fairness} (PoF). The PoF can take any positive value, although values closer to zero are preferable as they indicate lower loss of utility while enforcing fairness. The PoF can be defined mathematically as in \eqref{equ: POF}, and it will be compared for different fairness definitions. It is a continuous metric and independent of population size.
\begin{equation}
    POF = \frac{HC_{uti} - HC_{fair}}{ HC_{uti}},
    \label{equ: POF}
\end{equation}
 where $HC_{fair}$ can be the utilitarian, egalitarian, bounded or bargaining HC. Notice that under the utilitarian HC approach, the PoF is zero. Conversely, in the case of a complete loss of hosting capacity, the PoF reaches its upper bound of one.

In order to quantify the inequality, the Gini coefficient is used. For the HC problem formulation, it can be calculated as in \eqref{equ: Gini}. The Gini coefficient always takes a value between $0$ and $1$, where $0$ is the most fair situation and $1$ is unfair.
\begin{equation}
    G = \frac{1}{2 \bar{P}_{DG} n^2} \sum_{i,j} |P_{DG,i} - P_{DG,j}|
    \label{equ: Gini}
\end{equation}
% \vspace{-10pt}

\pagebreak
% \vspace{-10pt}
\section{Numerical Results}
\label{chap4: numerical results}

\begin{figure}[!htbp]   % !t means top of page
    \centering
    \subfloat[Network 5 Feeder 1: RF1]{\includegraphics[width=0.45\linewidth]{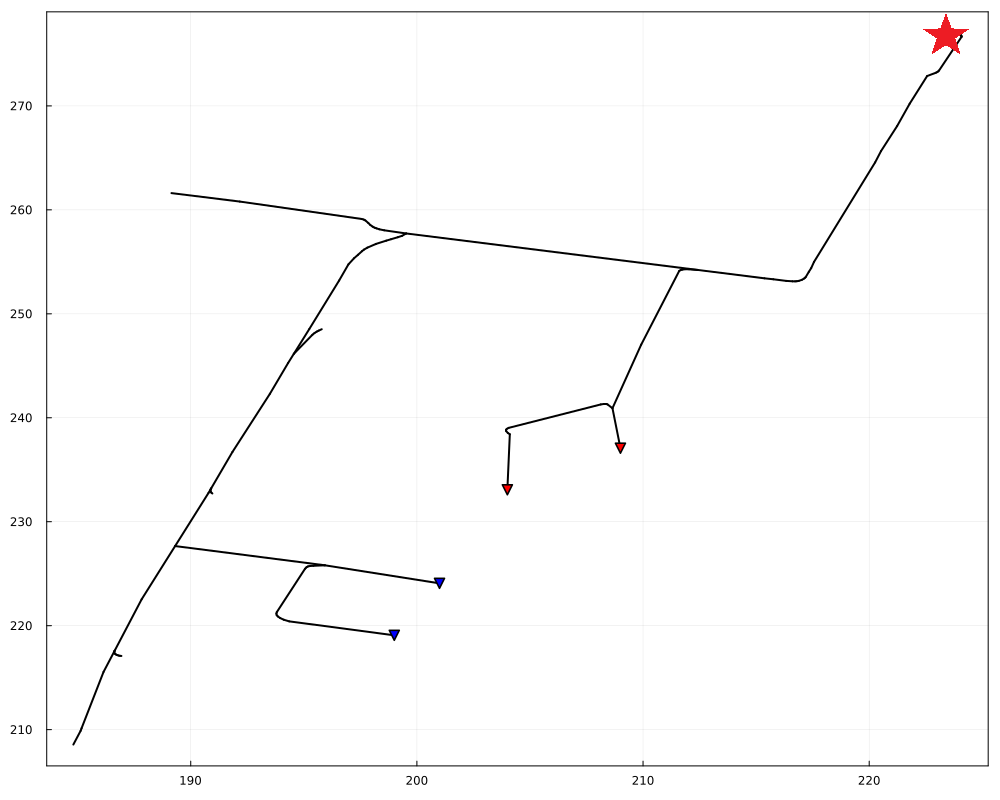}
    \label{fig:net5_feed1}}
    \hfil
        \subfloat[Network 2 Feeder 1: RF2]{\includegraphics[width=0.45\linewidth]{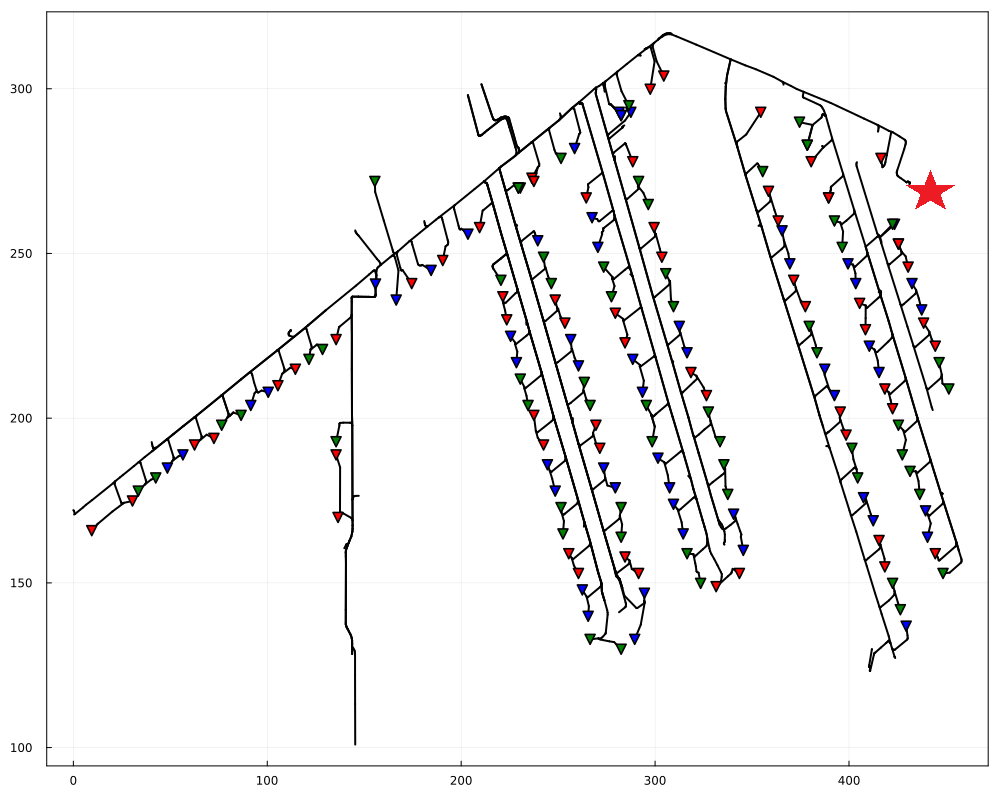}
    \label{fig:net2_feed1}}
    \hfil
    \subfloat[Network 5 Feeder 4: RF3]{\includegraphics[width=0.45\linewidth]{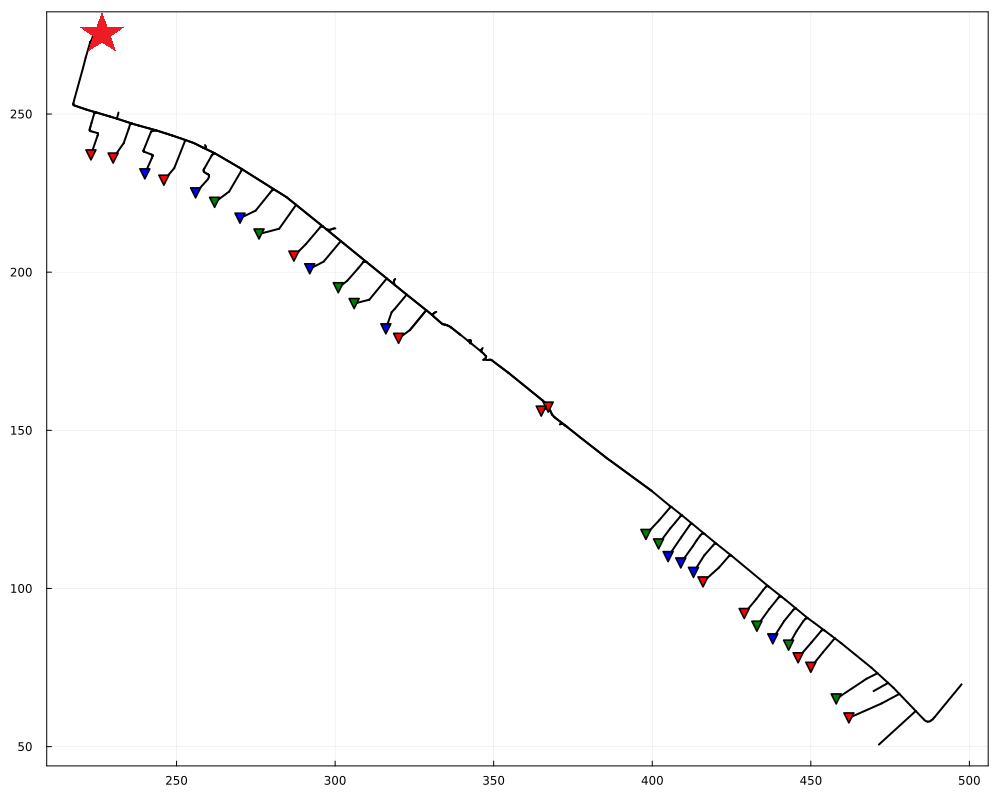}
    \label{fig:net5_feed4}}
    \hfil
    \subfloat[Network 2 Feeder 5: RF4]{\includegraphics[width=0.45\linewidth]{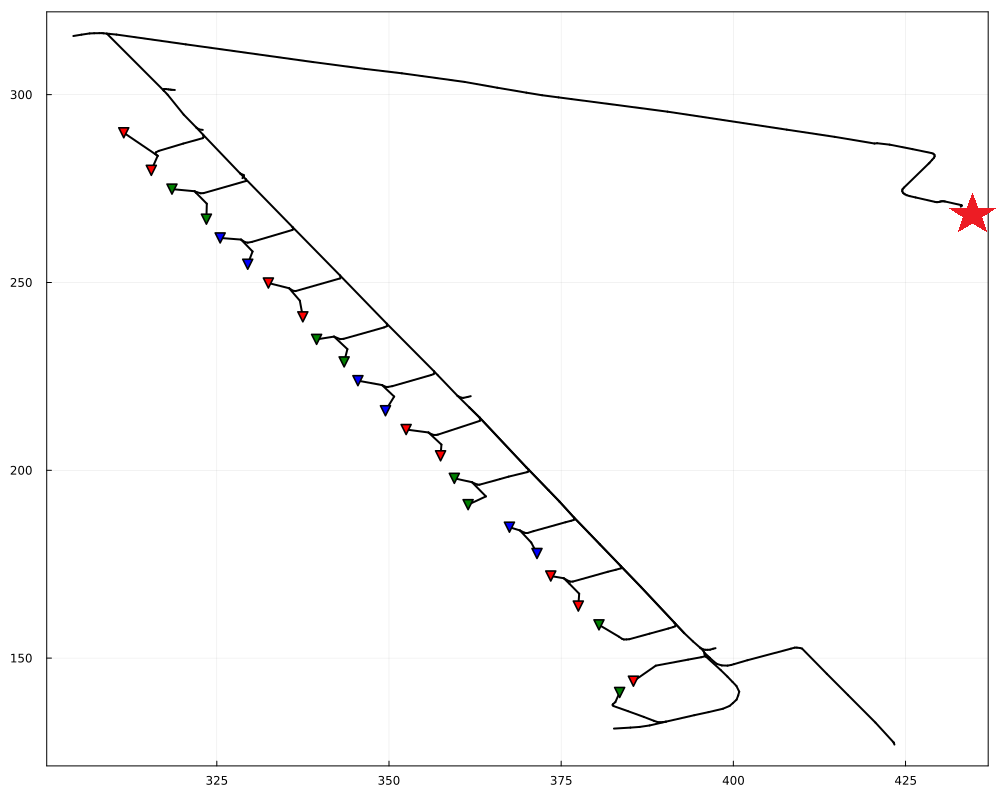}
    \label{fig:net2_feed5}}
    % \vspace{-7pt}
    \caption{\small{Topology of the four low-voltage representative feeders (RFs) used in the study \cite{hashmiidentifying}.}}
    \label{fig:4 feeders}
\end{figure}

% \subsection{Test cases}
The proposed approach is applied to four LV feeders sourced from the Electricity North West Limited (ENWL) dataset, the distribution network operator for the North West of England \cite{espinosa2015dissemination}. These feeders were selected to represent a range of network sizes, based on the number of loads and the length of the feeders. Specifically, RF1 is classified as small, RF3 and RF4 represent medium-sized feeders, and RF2 is considered large. The topology of the selected feeders is illustrated in Figure \ref{fig:4 feeders} and their key characteristics are summarised in Table \ref{tab: feeders features}.
The DG-HC optimisation problem is nonconvex, and we utilise IPOPT solver for identifying fairness-aware DG-HC values.

\begin{table}[h]
    \centering
    \scriptsize
    \caption{\small{Representative Feeder (RF) features \cite{hashmiidentifying}.}}
    % \vspace{-6pt}
    \begin{tabular}{c|c|c|c|c}
        \hline
        % Feeder & \makecell{Network 5 \\ Feeder 1} & \makecell{Network 2 \\ Feeder 1} & \makecell{Network 5 \\ Feeder 4} & \makecell{Network 2 \\ Feeder 5} \\
        % \hline
        id & RF1 & RF2 & RF3 & RF4 \\
        \hline
        feeder length ($km$) & 0.164 & 5.206  & 1.013 & 0.735 \\
        feeder resistance ($\Omega$)  & 0.219 & 4.617 & 0.907 & 0.376 \\
        feeder reactance ($\Omega$) & 0.014 & 0.416 & 0.081 & 0.055  \\
        R/X  & 14.100 & 10.479 & 12.971 & 12.286\\
        impedance ($\Omega$) & 0.22 & 4.64 & 0.91 & 0.38 \\
        number of loads & 4 & 175 & 30 & 23 \\
        number of buses & 12 & 359 & 65 & 52\\
        area ($m^2$) & 1280.1 & 67695.6 & 10291.7 & 14434.7 \\
        \hline
    \end{tabular}
    \label{tab: feeders features}
\end{table}
% \vspace{-15pt}
\subsection{Case 1: Assessing definitions of fairness}
The four feeders are compared to evaluate the effect of different fairness implementations. The analysis begins with a visual comparison of HC value under utilitarian and egalitarian fairness. This is followed by an examination of the corresponding Pareto frontiers for each feeder.
% \vspace{-10pt}
\begin{figure}[h]
    \centering
    \includegraphics[width=0.627\linewidth]{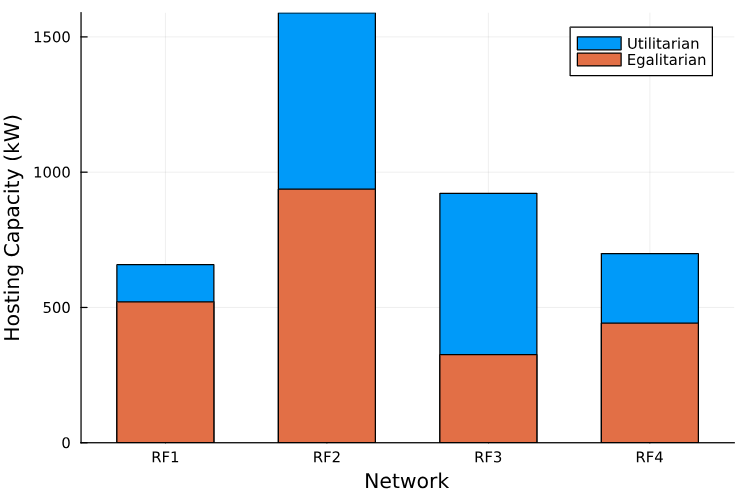}
    % \vspace{-6pt}
    \caption{\small{Egalitarian and utilitarian HC for the 4 feeders (RF1: small size, RF2: large size, RF3: medium size with less branching, RF4: medium size with more branching).}}
    \label{fig: ega and uti for 4 networks}
\end{figure}

From Figure \ref{fig: ega and uti for 4 networks}, it can be concluded that the egalitarian scenario is inherently more restrictive than the utilitarian one, since the voltage constraint at the most limiting bus creates a global bottleneck. Once this limit is reached, no additional DG can be integrated, regardless of the capacity available at other buses. This constraint uniformly affects all four feeders under consideration, resulting in lower HC values. However, the variation in HC reduction among feeders is attributed to differences in their topologies, as discussed in detail in case $2$.

\begin{figure}[!t]   % !t means top of page
    \centering
    \subfloat[RF1]{\includegraphics[width=0.470\linewidth]{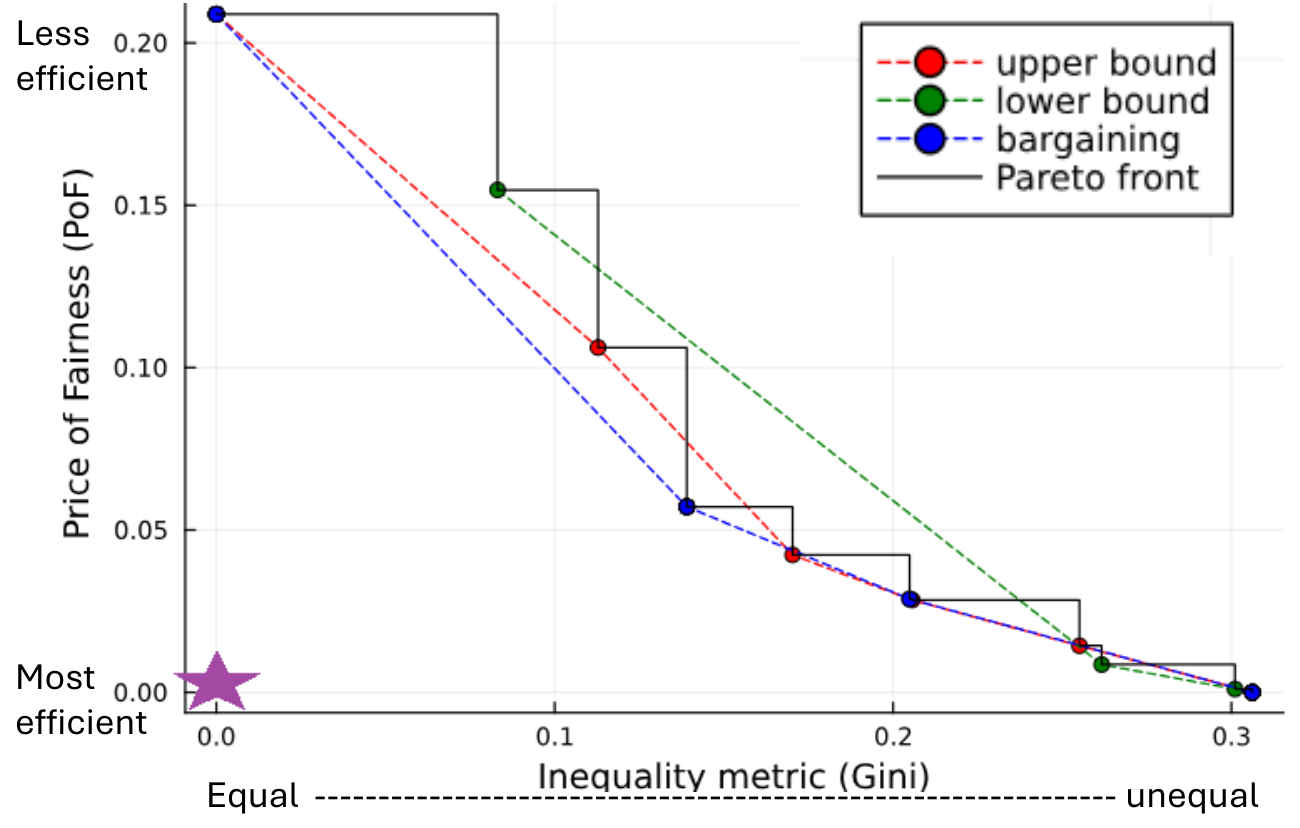}
    \label{fig: pareto curve 51}}
    \hfil
    \subfloat[RF2]{\includegraphics[width=0.470\linewidth]{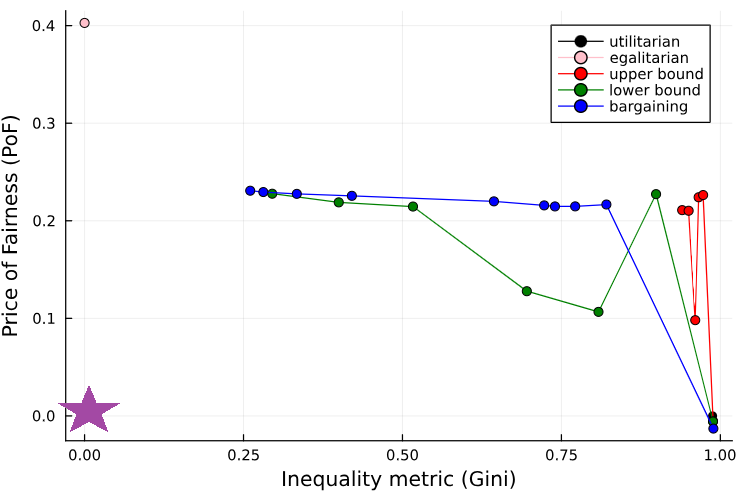}
    \label{fig: pareto curve 21}}
    \hfil
    \subfloat[RF3]{\includegraphics[width=0.470\linewidth]{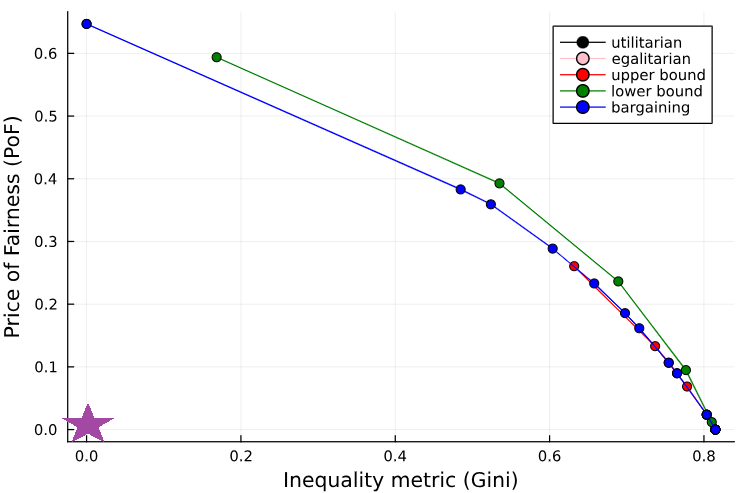}
    \label{fig: pareto curve 54}}
    \hfil
    \subfloat[RF4]{\includegraphics[width=0.470\linewidth]{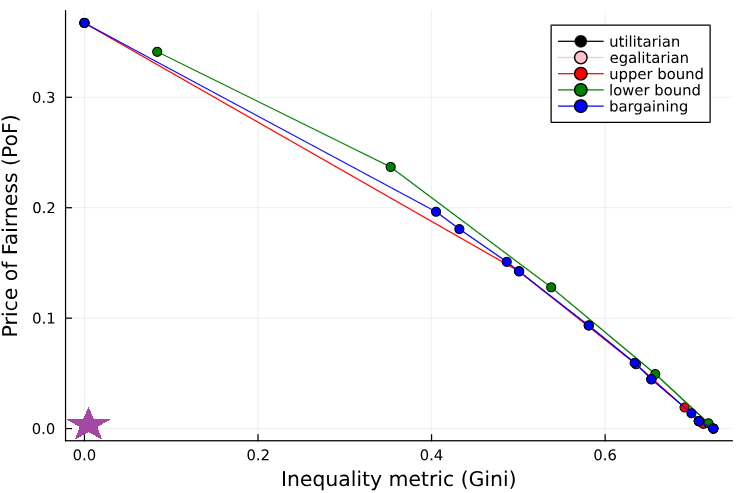}
    \label{fig: pareto curve 25}}
    % \vspace{-8pt}
    \caption{\small{Pareto frontiers for different fairness definitions (utilitarian, egalitarian, bargaining, upper and lower bounded) for the 4 RFs.}}
    \label{fig: pareto curves}
\end{figure}

The Pareto plots of the different definitions are discussed below for all feeders and can be seen in Figure \ref{fig: pareto curves}. The Pareto frontiers illustrate the results for the different fairness definitions across the networks, with the utilitarian and egalitarian solutions occupying the two extremes. The optimal solution for the Pareto frontiers is situated on the bottom left-hand corner, since inequality increases along the \textit{x}-axis and efficiency decreases along the \textit{y}-axis.
For the medium and small feeders, the results of the different fairness definitions lie relatively close to each other, but the bargaining and upper bounded solutions consistently position themselves closer to the optimal point. Notably, some lower-bound solutions lie beyond the Pareto frontier for the medium feeders (especially RF3), meaning they perform worse in terms of both PoF and Gini coefficient.

A notable deviation is observed in the behaviour of RF2, the large feeder. Unlike the other feeders, RF2 exhibits distinct dynamics due to its size and structural complexity. Although it is technically a single feeder, it comprises multiple branches, likely corresponding to different streets, see Fig. \ref{fig:net2_feed1}. This complex topology explains the deviations regarding the upper bounded and bargaining fairness definition. The upper bound cluster is explained with the presence of an outlier, a load having the ability to produce $1000$ kW. As this value is extreme and the upper bound reduces linearly, the different HC solutions do not vary greatly and fairness is not increased. Next, the bargaining solution gives almost constant values of the PoF for increasing the Gini coefficient. This minimal efficiency loss observed during continued improvements in fairness suggests that the feeder's complexity and the large number of loads enable flexible redistribution of DG capacity. Lastly, the lower bounded fairness features a sudden increase in PoF as inequality increases. This outlier is difficult to explain, since a decrease in the lower bound is expected to lead to an increase in efficiency as the feasible solution space becomes greater.
The optimal values of negotiated fairness HC solutions applying bargaining, upper and lower bounded solutions depend on the selection of parameters K, $\beta, ~\alpha$. These parameters can be identified using \textit{knee point identification} of the Pareto front curves shown in Fig. \ref{fig: pareto curves}.

\pagebreak

\subsection{Case 2: Network topology sensitivity}
\label{case 2}
\begin{figure}[h]
    \centering
    \includegraphics[width=0.7\linewidth]{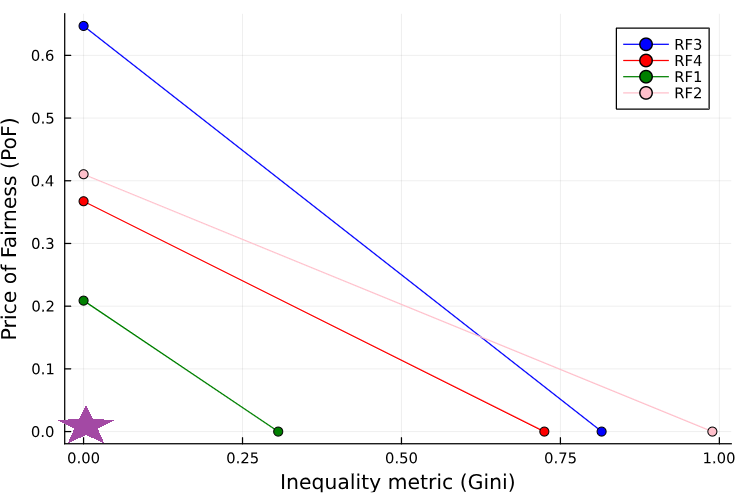}
    % \vspace{-7pt}
    \caption{\small{PoF vs Gini coefficient for utilitarian and egalitarian solution of all feeders with the star symbol representing the optimal solution.}}
    \label{fig: PoF vs Gini for the 4 feeders}
\end{figure}

The influence of feeder size and topology on HC is evaluated by comparing the four feeders. The analysis focuses on the egalitarian and utilitarian allocation strategies, as illustrated in Figure \ref{fig: PoF vs Gini for the 4 feeders} and detailed in Table \ref{tab: ega and uti data for 4 networks2}.
% \vspace{-10pt}
% \vspace{-15pt}
\begin{table}[h]
    \centering
    \caption{\small{Utilitarian and egalitarian HC for the 4 feeders with the Gini coefficient evaluating the inequality of the utilitarian solution and the PoF assessing the efficiency of the egalitarian solution.}}
    \begin{tabular}{|c|c|c|c|c|}
        \hline
         RF & $HC_{uti}$ [$kW$] & $Gini_{uti}$ & $HC_{ega}$ [$kW$] & $PoF_{ega}$  \\
         \hline
         $1$ & 658  & 0.3062 & 520 & 0.21 \\
         $2$ & 1575 & 0.9876 & 937 & 0.41  \\
         $3$ & 922  & 0.8148 & 325 & 0.65 \\
         $4$ & 699  & 0.7247 & 442 & 0.37 \\
         \hline
    \end{tabular}
    \label{tab: ega and uti data for 4 networks2}
\end{table}

Fig. \ref{fig: PoF vs Gini for the 4 feeders} illustrates the relationship between the PoF and the Gini coefficient across different feeder sizes. For large feeders, such as RF2, the utilitarian solution is the least fair. However, the relatively gentle slope of the curve suggests that achieving a fairer allocation involves only a moderate sacrifice in utility. Among the medium-sized feeders, RF3, which has low branching, shows a steep trade-off curve, indicating a significant reduction in utility to gain fairness. In contrast, RF4, a highly branched medium feeder, displays more stable behaviour with a moderate trade-off between efficiency and fairness. The small RF1 shows a trade-off slope similar to that of RF4 but starts from a lower Gini coefficient under the utilitarian solution, suggesting an inherently fairer configuration. %These observations are examined in greater detail in the following paragraphs.

Differences in utilitarian HC across feeders result from variations in both size and topology. Although it might be intuitive to expect HC to scale with feeder size, measured in total length and number of connected loads (see Table \ref{tab: feeders features}), this relationship does not consistently hold. For example, the HC difference between small and medium feeders is marginal, only $5.8\%$, despite the medium feeder serving nearly six times as many loads. This limited gain can be attributed to greater voltage drops in larger feeders, which occur as feeder length increases. Additionally, this example highlights the fact that feeders are not always optimally designed. In this case, the small feeder may be overdimensioned, as the individual DG capacities cannot be achieved in practice, even under the egalitarian solution. 

Furthermore, feeders of a similar size may respond differently to fairness constraints due to topological differences. For example, RF3 and RF4 have comparable lengths and number of loads, but react differently to the egalitarian implementation. From Table \ref{tab: ega and uti data for 4 networks2} it can be deduced that RF3 experiences a 65$\%$ reduction in HC under egalitarian allocation, whereas RF4 incurs only a 37$\%$ loss. This divergence arises from their structural configurations. Both feeders have a transformer located at one end, but RF4 has numerous short lateral branches, whereas RF3 is predominantly linear, with sequentially connected buses and limited branching. These topological differences affect the voltage drops and, consequently, the feeder's capacity to accommodate distributed generation under different allocation schemes. 
% \vspace{-15pt}

\begin{figure}[h]
    \centering
    \includegraphics[width=0.76\linewidth]{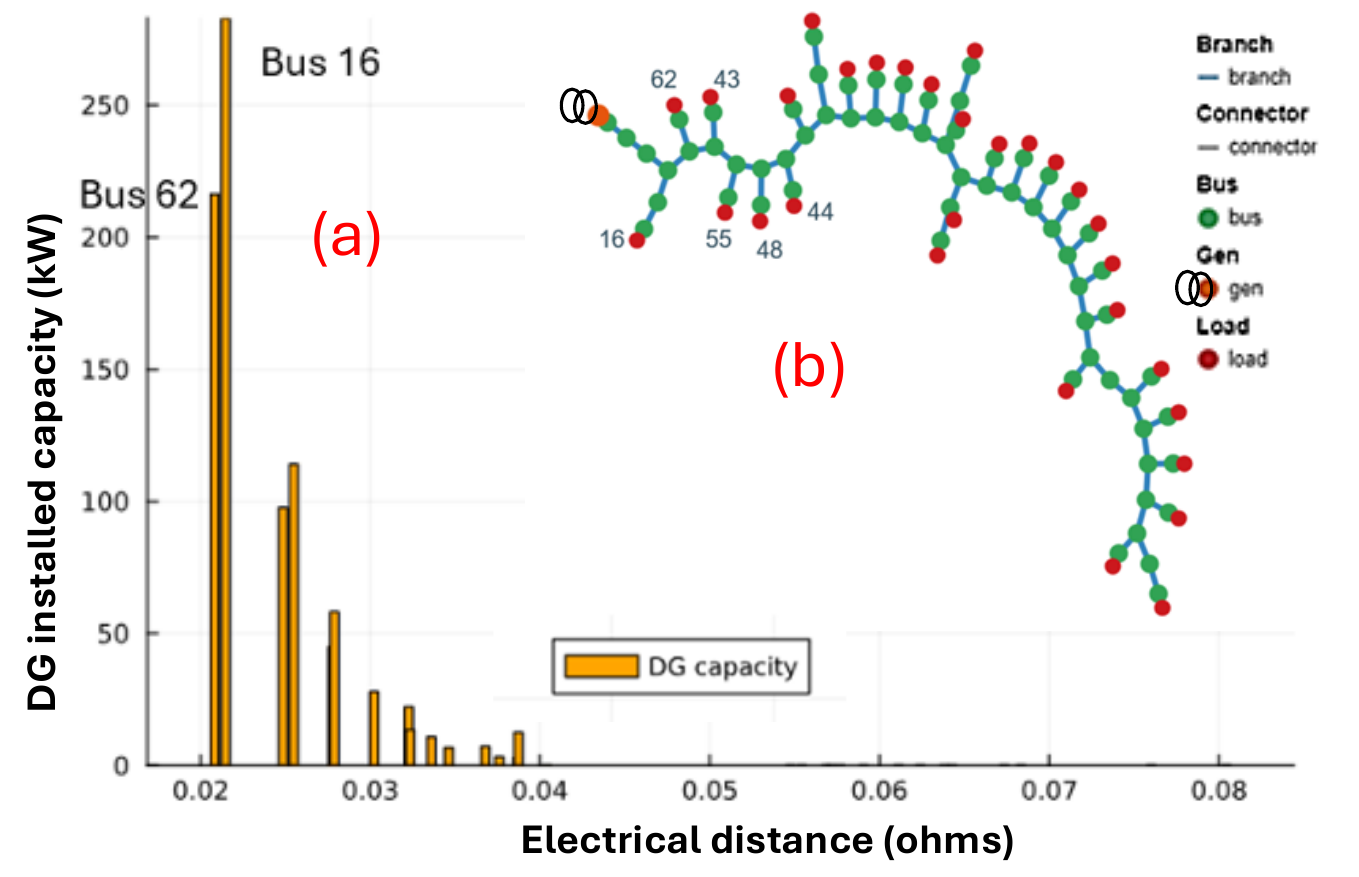}
    % \vspace{-7pt}
    \caption{\small{Utilitarian DG HC solution RF3 plotted with the electrical distance from the substation.}}
    \label{fig: RF3_dg_utili}
\end{figure}

Fig. \ref{fig: RF3_dg_utili} shows the nodal DG installed capacity as an outcome of solving the DG HC problem. Note that DG installation, if not bounded, is preferred close to the substation. Of course, this distribution of DGs leads to increased DG HC, however, the disparity metric, i.e., Gini index, exceeds 0.8, see Fig. \ref{fig: pareto curve 54}.

\pagebreak

\section{Conclusion}
\label{chap5: conclusion}
This study investigated how fairness considerations influence HC in various LV feeder configurations. An optimisation-based framework was developed to maximise HC while incorporating fairness, either by introducing additional constraints or by modifying the objective function. Multiple notions of fairness definitions were implemented to explore different fair capacity allocation approaches.
Pareto plots were used to evaluate and compare the performance of the various fairness definitions, illustrating the trade-offs between efficiency and fairness. The primary performance indicators used in this analysis were the Price of Fairness (PoF) and the Gini coefficient. Of the different fairness formulations, the bargaining and upper-bounded approaches produced the most consistent and robust results, balancing efficiency and fairness effectively across feeders of different sizes and topologies. In contrast, the lower-bound fairness approach produced less consistent outcomes overall.

Both feeder topology and size have a significant impact on HC and fairness. Feeder size primarily affects HC for small to medium-sized feeders, demonstrating substantially more favourable results than large feeders. In contrast, feeder topology seems to have a more pronounced impact on fairness outcomes. Branched topologies, which minimise the presence of long, shared feeder branches, tend to provide the most equitable distribution of DG capacity without losing too much efficiency.

In future work, stochasticity and phase imbalance should be explored to assess their impact on HC, fairness in DG allocation, and the extent to which they may constrain it. 
Finally, extending the notions of fairness for other types of distributed energy resources, such as EVs, needs further exploration.
% Additionally, expanding the analysis to include a broader range of feeder types would validate the current results and facilitate the identification of design principles for optimising LV feeder performance.

% \section*{Acknowledgement}
% This work is supported by the Flemish Government and Flanders Innovation \& Entrepreneurship (VLAIO) through the Flux50 project IMPROcap (HBC 2022.0733) {and KU Leuven funded C2 project FlexIQ}.

\pagebreak

\bibliographystyle{IEEEtran}
\bibliography{reference}

\end{document}